\documentstyle[epsfig,aps,twocolumn,prl,floats,tighten]{revtex}
\newcommand{\ires}       {$\rm^{1}$}
\newcommand{\gsi}        {$\rm^{2}$}
\newcommand{\heidelberg} {$\rm^{3}$}
\newcommand{\korea}      {$\rm^{4}$}
\newcommand{\clt}        {$\rm^{5}$}
\newcommand{\bucarest}   {$\rm^{6}$}
\newcommand{\zagreb}     {$\rm^{7}$}
\newcommand{\itep}       {$\rm^{8}$}
\newcommand{\budapest}   {$\rm^{9}$}
\newcommand{\warsaw}     {$\rm^{10}$}
\newcommand{\dresde}     {$\rm^{11}$}
\newcommand{\kur}        {$\rm^{12}$}
\begin{document}
\draft

\title{Isospin-tracing:  A probe of non-equilibrium
in central heavy-ion collisions}

\author{
F.\,Rami\ires,
Y.~Leifels\gsi$^{\rm ,}$\heidelberg,
B.\,de Schauenburg\ires,
A.~Gobbi\gsi,
B.\,Hong\korea,
J.P.\,Alard\clt,
A.\,Andronic\bucarest,
R.~Averbeck\gsi,
V.\,Barret\clt,
Z.\,Basrak\zagreb,
N.\,Bastid\clt,
I.\,Belyaev\itep,
A.\,Bendarag\clt,
G.\,Berek\budapest,
R.\,\v{C}aplar\zagreb,
N.\,Cindro\zagreb,
P.~Crochet\clt,
A.~Devismes\gsi,
P.\,Dupieux\clt,
M.\,D\v{z}elalija\zagreb,
M.\,Eskef\heidelberg,
C.~Finck\gsi,
Z.\,Fodor\budapest,
H.~Folger\gsi,
L.\,Fraysse\clt,
A.\,Genoux-Lubain\clt,
Y.\,Grigorian\gsi,
Y.\,Grishkin\itep,
N.~Herrmann\gsi$^{\rm ,}$\heidelberg,
K.D.~Hildenbrand\gsi, 
J.\,Kecskemeti\budapest,
Y.J.\,Kim\korea,
P.\,Koczon\gsi,
M.\,Kirejczyk\warsaw,
M.\,Korolija\zagreb,
R.\,Kotte\dresde,
M.\,Kowalczyk\warsaw,
T.~Kress\gsi,
R.~Kutsche\gsi,
A.\,Lebedev\itep,
K.S.\,Lee\korea,              
V.\,Manko\kur,
H.\,Merlitz\heidelberg,
S.\,Mohren\heidelberg,
D.\,Moisa\bucarest,
J.\,M\"osner\dresde,
W.\,Neubert\dresde,
A.\,Nianine\kur,
D.\,Pelte\heidelberg,
M.\,Petrovici\bucarest,
C.\,Pinkenburg\gsi,
C.\,Plettner\dresde,
W.~Reisdorf\gsi,
J.~Ritman\gsi,
D.~Sch\"ull\gsi, 
Z.\,Seres\budapest,
B.\,Sikora\warsaw,
K.S.\,Sim\korea,
V.\,Simion\bucarest,
K.\,Siwek-Wilczy\'nska\warsaw,
A.\,Somov\itep,
M.R.\,Stockmeier\heidelberg,
G.\,Stoicea\bucarest,
M.\,Vasiliev\kur,
P.\,Wagner\ires,
K.~Wi\'{s}niewski\gsi,
D.\,Wohlfarth\dresde,
J.T.\,Yang\korea,
I.\,Yushmanov\kur,
A.\,Zhilin\itep\\
the FOPI Collaboration 
}

\address{
\ires~Institut de Recherches Subatomiques, IN2P3-CNRS, Universit\'e
Louis Pasteur, Strasbourg, France \\
\gsi~Gesellschaft f\"ur Schwerionenforschung, Darmstadt, Germany\\
\heidelberg~Physikalisches Institut der Universit\"at Heidelberg, 
Heidelberg, Germany\\
\korea~Korea University, Seoul, South Korea\\
\clt~Laboratoire de Physique Corpusculaire, IN2P3/CNRS,
and Universit\'{e} Blaise Pascal, Clermont-Ferrand, France\\
\bucarest~National Institute for Nuclear Physics and Engineering, 
Bucharest, Romania\\
\zagreb~Rudjer Boskovic Institute, Zagreb, Croatia\\ 
\itep~Institute for Theoretical and Experimental Physics, Moscow, Russia\\
\budapest~Central Research Institute for Physics, Budapest, Hungary\\
\warsaw~Institute of Experimental Physics, Warsaw University, Poland\\
\dresde~Forschungszentrum Rossendorf, Dresden, Germany\\
\kur~Kurchatov Institute, Moscow, Russia \\
}

\maketitle

\begin{abstract}
Four different combinations of $^{96}_{44}$Ru and $^{96}_{40}$Zr
nuclei, both as projectile and target, were investigated at the same
bombarding energy of 400$A$ MeV using a $4 \pi$ detector.
The degree of isospin mixing between projectile and target nucleons is
mapped across a large portion of the phase space using two
different isospin-tracer observables, the
number of measured protons and the ${\rm t}/^{3}{\rm He}$ yield ratio.
The experimental results show that the global equilibrium is not reached  
even in the most central collisions. Quantitative measures  
of stopping and mixing  are extracted from the data. 
They are found to exhibit a quite strong sensitivity
to the in-medium (n,n) cross section used in microscopic 
transport calculations.
\end{abstract}


Central heavy-ion collisions represent  a valuable tool for studies of hot 
and dense nuclear matter where one hopes to infer valuable information
on the nuclear equation of state ({\small EOS})  and on modifications of
hadrons in the nuclear medium.
It is still an open
question whether the widely applied, at least local if not global, equilibrium
assumption is valid in such reactions~\cite{sto 86,ber 84}, or whether 
significant
non-equilibrium effects rather require the application of more elaborated
non-equilibrium dynamical models~\cite{ber 84,aic 91,fel 90}.
The issue of equilibration is expected to be influenced by
in-medium effects (such as Pauli blocking, Fermi motion) on the 'hard'
scattering processes, by early 'soft' deflections in the 
momentum-dependent mean fields, and by finite-size (corona) effects.
An understanding of all these effects is a prerequisite for a quantitative 
extraction of the {\small EOS} from nucleus-nucleus collisions.  
 
Experimental observations 
of non-equilibrium 
in relativistic heavy-ion collisions
were concentrated,
up to now, on the measurement of the momentum distribution of the products
emerging from a mid-rapidity "source" of symmetric colliding systems.
Observables of interest were
the width of rapidity distributions~\cite{aic 91,hon 98} 
or the overall shape of the source~\cite{jeo 94,rei 97}.
The sensitivity of such observables is however reduced by effects
like rescattering during the late phase of expansion.

In order to extract, in a model independent approach, direct experimental
information on non-equilibrium we have designed a new type of
high precision measurement which makes use of the isospin (${N/Z}$) 
degree of freedom.
The (${N/Z}$) equilibration has been investigated before at low bombarding
energies in fusion-like reactions~\cite{yen 94,joh 96,joh 97}.
As it will be shown in some details later, (${N/Z}$) 
can be used as a tracer in order to attribute the measured nucleons 
(on average) either to the target or to the projectile nucleons. 
It is therefore possible to extract rapidity-density distributions  
separately for projectile and target nucleons. This gives access to 
new more sensitive observables, like stopping and mixing,
of the  early equilibration process.    

The experiment was carried-out using reactions between equal mass nuclei 
$A=96$, at an incident energy of 400$A$ MeV. Isotopes of 
Ru and Zr were taken as projectile and target making use of all the four 
combinations: Ru+Ru, Ru+Zr, Zr+Ru and Zr+Zr. This choice of isotopes
takes advantage of an almost unique possibility offered by the periodic
table of stable isotopes, while searching for two isobars of the largest
possible ${N/Z}$ difference (the ${N/Z}$ ratio is equal to
$1.18$ and $1.40$ for $^{96}_{44}$Ru and $^{96}_{40}$Zr, respectively)
which can be used both as projectile and target.      
The bombarding energy was chosen at the minimum of the 
(n,n) cross section where the relative motion is 
significantly larger than the Fermi motion, but sufficiently low
to avoid inelastic (n,n) channels, while the (n,n) angular distribution is 
almost isotropic.   

The experiment was performed at {\small SIS/ESR}-Darmstadt using the
{\small FOPI} apparatus~\cite{gob 93,rit 95}.
The reaction products were detected at forward angles by the
{\small TOF}-wall and the {\small HELITRON} drift chamber
and at backward angles by the Central Drift Chamber ({\small CDC}).
Particles of interest here were: i) tritons and $^3$He emitted at 
laboratory angles ${10}^{\circ} < \theta_{\rm lab} < {28}^{\circ}$,
identified by combining the {\small TOF}-wall and the {\small HELITRON}
and ii) protons and deuterons emitted at backward angles
(${34}^{\circ} < \theta_{\rm lab} < {145}^{\circ}$), identified by the 
{\small CDC}.
These particles constitute a large fraction of the
emitted charge, about $40\%$, $30\%$, $10\%$ and $10\%$ for 
p, d, t and $^3$He, respectively.

\vspace{-0.7cm}

\begin{figure}[htp]
\hspace{0.2cm}\epsfig{file=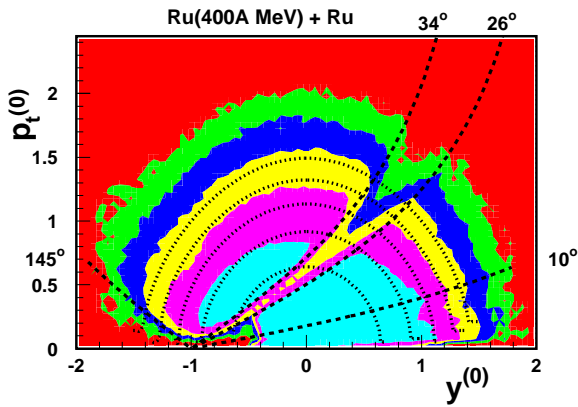,height=6.cm}
\caption{
Invariant cross section in the
($y^{(0)}$, $p_t^{(0)}$) plane of $Z=1$ particles measured
in the CDC and the TOF-wall, under the $E_{\rm RAT} > 1.4$ centrality cut.
$y^{(0)}$ is the normalized rapidity, {\rm i.e.} the particle
{\rm c.m.} rapidity divided by the projectile rapidity in the {\rm c.m.}
system. $p_t^{(0)}$ denotes the normalized transverse momentum, {\rm i.e.}
the particle transverse momentum per nucleon
divided by the {\rm c.m.} projectile momentum per nucleon.
The grey levels correspond to different logarithmic cuts
in the invariant  cross section.
The dashed lines show the acceptance of the CDC
(${34}^{\circ} < \theta_{\rm lab} < {145}^{\circ}$) and the HELITRON
(${10}^{\circ} < \theta_{\rm lab} < {26}^{\circ}$) detector components.
The circles correspond to constant {\rm c.m.} energies per nucleon:
from the inner to the outer, 40, 80, 120, 160 and 200$A$ MeV.
}
\label{pty}
\end{figure}

Central collisions were selected requiring a large ratio of total
transverse to total longitudinal energy,  
$E_{\rm RAT}$ \cite{jeo 94,rei 97}.
$E_{\rm RAT}$ is determined event wise
by including \hbox{all} detected particles: 
$E_{\rm RAT}={\sum_{\rm i=1}^{\rm M}E_{\rm t}^{\rm i}} / 
{\sum_{\rm i=1}^{\rm M}E_{\rm l}^{\rm i}}$, 
where $E_{\rm t}^{\rm i}$ and $E_{\rm l}^{\rm i}$ are
the transverse and longitudinal kinetic energies of the particle ${\rm i}$ 
and ${\rm M}$ is the number of detected particles. The $E_{\rm RAT} > 1.4$ 
cut used here \hbox{corresponds} to about $1.5 \%$ of the total reaction
cross section and to a geometrical impact parameter
$b_{\rm geom} \le 1.3\;$fm in a sharp-cut-off approximation.

Fig. 1 shows that the events selected under the $E_{\rm RAT} > 1.4$ condition
exhibit, in a representation of rapidity versus transverse momentum,
a nearly isotropic source for the observed hydrogen ($Z=1$) products. 
This is in 
agreement with previous observations~\cite{hon 98,jeo 94,rei 97,lis 95}.

The (${N/Z}$)-tracer  method is based on the following idea:
let us assume that we are observing  the final number of protons, $Z$
in a given cell of the momentum space. 
The expected yield $Z^{\rm Ru}$ measured   for the
Ru+Ru reaction is higher than $Z^{\rm Zr}$ of the Zr+Zr reaction since 
Ru has $44$ protons    as opposed to $40$ for Zr.   
Such measurements using identical projectile and target deliver  
calibration values  $Z^{\rm Ru}$ and $Z^{\rm Zr}$ for each observed
cell. In the case of a mixed reaction, Ru+Zr or Zr+Ru, the measured
proton yield $Z$ takes values intermediate between the calibration
values ($Z^{\rm Ru}$, $Z^{\rm Zr}$).
If {\rm e.g.} $Z$ is close to  $Z^{\rm Ru}$ in a Ru+Zr reaction,
means that the cell is populated predominantly from nucleons of the 
Ru-projectile while if it is  close to  $Z^{\rm Zr}$ it is mostly
populated from nucleons of the Zr-target.
In this way it is possible to trace back the relative abundance of
target to projectile nucleons contributing to a given cell. 
If the mixed value $Z$ varies linearly between the two extremes of the 
calibration values  ($Z^{\rm Ru}$, $Z^{\rm Zr}$) as a function of the 
relative abundance of target versus projectile nucleons, such an 
abundance can be most easily derived.
As it will be shown this is indeed the case for protons but it is not 
the case if we use as a (${N/Z}$)-tracer variable not the 
number of protons  but 
{\rm e.g.} the relative tritium to  $^3$He abundance in the cell. 
The ${\rm t}/^3{\rm He}$-ratio varies non-linearly between the 
two extreme of the
calibration reactions, but this can be taken  into account   and, 
as it will be shown, consistent results  are obtained with  different
(${N/Z}$)-tracer variables.

The following definition for the relative abundancy of the 
projectile-target nucleons has been adopted:
                 
\begin{equation}
R_{\rm Z} \quad = \quad {{2 \times Z - Z^{\rm Zr} - Z^{\rm Ru}}  \over
{Z^{\rm Zr} - Z^{\rm Ru}}}
\end{equation}

\noindent where $R_{\rm Z}$ takes $+1$ for Zr+Zr and $-1$ for Ru+Ru.
In the case of full equilibrium in a mixed reaction,
$R_{\rm Z}$
would  be equal  to $0$ everywhere independently of the
location of the cell (for the case of a linear dependence of $Z$).

Several important advantages of the method can be mentioned:
(i) the four reaction combinations are investigated, under identical
experimental conditions so that the ratios are insensitive to systematic 
uncertainties due to the apparatus. 
The errors are essentially of statistical nature
and profit from the high yield in the considered cell.
(ii) the mixed reaction Zr+Ru is the same as Ru+Zr except that target and 
projectile are inverted: this allows forward-backward cross-checks of the
apparatus which in addition can also be obtained from the symmetric Ru+Ru
and Zr+Zr reactions.
(iii) using the four reactions the full information needed can be obtained
by measuring only within the center-of-mass ({\rm c.m.}) backward or
only the {\rm c.m.} forward hemisphere.

\vspace{-0.4cm}

\begin{figure}[htp]
\hspace{0.3cm}\epsfig{file=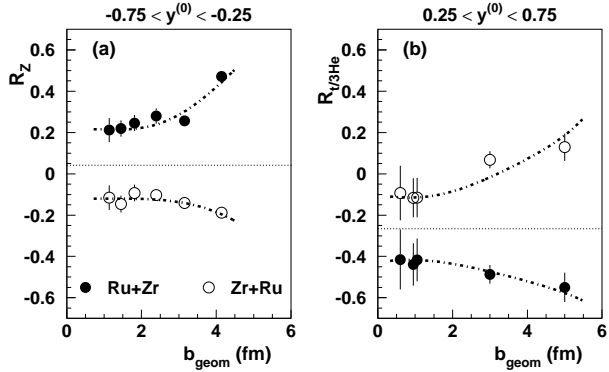,height=5.2cm}
\caption{
Two isospin-tracer observables as a function of the geometrical impact 
parameter: $R_{\rm Z}$ (panel a)
and $R_{{\rm t}/^{3}{\rm He}}$ (panel b).
They are determined in the {\rm c.m.} rapidity range
$-0.75 \le {y^{(0)}} \le -0.25$  and 
$0.25 \le {y^{(0)}} \le 0.75$, respectively.
The horizontal dotted lines correspond to the values
($R_{\rm Z} = 0.04$ and $R_{{\rm t}/^{3}{\rm He}}=-0.27$) at 
halfway between the
Ru+Zr and Zr+Ru experimental points in the
$b_{\rm geom} \le 1.5 \;$fm region.
The dotted-dashed lines are just to guide the eye.
}
\label{v1}
\end{figure}

The results to be presented here concentrate on two  
tracer-observables: (A) the sum ($Z$) of the number of detected
free protons plus the number of protons detected within the deuterons.
(B) the relative abundance of tritium to $^3$He (${{\rm t}/^{3}{\rm He}}$).
Observable (A) was measured in the backward {\rm c.m.} hemisphere
using the {\small CDC}: this has the advantage to minimize background 
originating from the oxygen content of the  (ZrO$_2$) target. 
Observable (B) was measured in the forward {\rm c.m.} hemisphere
using {\small HELITRON} and {\small TOF}-wall  
profiting from a good time-of-flight particle identification. 

Results on the centrality dependence of the abundancy ratios are presented in 
Fig. 2. The selected momentum cell is rather wide and integrates over a 
large rapidity bin. 
Panels (a) and (b) show results obtained from, 
proton yields and ${\rm t}/^3{\rm He}$-ratios, respectively. 
The quantity $R_{{\rm t}/^3{\rm He}}$ is defined in a similar way as
for $R_{\rm Z}$, using \hbox{equation (1)} with $Z$ being replaced by
the ${\rm t}/^{3}{\rm He}$ abundancy ratio.
The impact parameter 
$b_{\rm geom}$ is derived by integrating over the measured cross section
as a function of $E_{\rm RAT}$ or of the charged particle multiplicity.
Except for an off-set of $R_{{\rm t}/^3{\rm He}} = -0.27 \pm 0.07$ for the 
${\rm t}/^3{\rm He}$-ratio, both figures display the same trends. 
Such an off-set can be understood from a non-linear 
dependence of the  ${\rm t}/^3{\rm He}$ ratio as a function of isospin. 
We find that, for $N > Z$, an empirical dependence of the 
${\rm t}/^{3}{\rm He}$ ratio of
[$1 + 100 \times \left(\frac{N-Z}{N+Z}\right)^{2.5}$]
describes the known systematics (see Fig.35 in ref.~\cite{nag 81})
and satisfies the measured off-set of $-0.27$. 
The off-set on Fig. 2a is very small ($0.04 \pm 0.03$) and demonstrates
the linear dependence of variable (A), 
what is also confirmed in  Fig. 3a: at 
mid-rapidity  ($y^{(0)} =0$) the ratio $R_{\rm Z}$ is very
close to $0$, for  both mixed reactions as expected from symmetry.

The results of Fig. 2 are consistent with the picture of an
incomplete global equilibrium which persists up to the end of the collision,
an observation that holds even for the most central collisions,
for $b_{\rm geom} \rightarrow 0$.

\vspace{-0.7cm}
\begin{figure}[htp]
\hspace{0.2cm}\epsfig{file=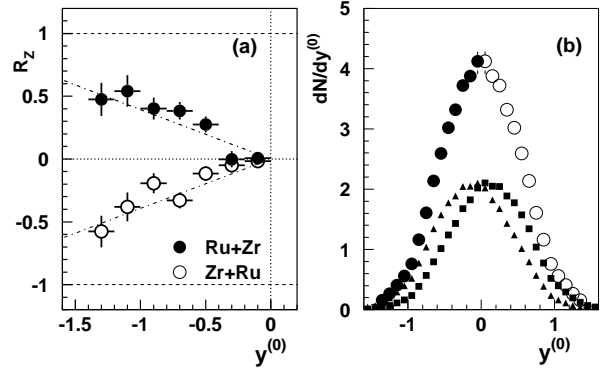,height=5.8cm}
\caption{
Panel (a) shows the tracer observable  $R_{\rm Z}$ as a function of the
normalized {\rm c.m.} rapidity for central collisions. 
The experimental results are shown for both
isospin asymmetric reactions Ru+Zr (full circles) and Zr+Ru (open circles).
The error bars correspond to statistical uncertainties.
The horizontal dashed lines indicate the values of $ R_{Z}= +1$ and
$-1$ corresponding to the isospin symmetric reactions
Zr+Zr and Ru+Ru, respectively. 
Panel (b) displays the experimental {\rm c.m.}
rapidity distribution (circles) of protons, free plus those detected 
within the deuterons, in 
central Ru+Ru collisions. Data points in the backward hemisphere
(full circles) were obtained from the CDC detector. 
Those in the forward region (open circles) were obtained
by assuming a backward/forward symmetry. 
The two other distributions have been obtained
by unfolding the overall distribution into 'projectile' (squares)
and 'target' (triangles) components. 
}
\label{dNdY}
\end{figure}

After the inspection of the centrality dependence of Fig. 2, we
show in Fig. 3 the rapidity dependence of $R_{\rm Z}$ for the
most central collisions. For this purpose we use the proton yield (A)
whose relatively high cross sections permit a fine subdivision
of the rapidity \hbox{bins.} Inverting projectile and target changes, as
expected, the sign of the $R_{\rm Z}$-values; besides that, both
results agree within errors. 
The dashed-dotted line ($\pm 0.393 \; y^{(0)}$) describes an 
average
of both measurements (being the same except for the sign) and can
be used to deconvolute the overall measured rapidity distribution
d$N/$d$y^{(0)}$ (Fig. 3b) for Ru+Ru into 
{\it separated rapidity distributions 
for the projectile- and for the target-nucleons}.
For each rapidity bin, the number of projectile (target) nucleons
was obtained as 
$N_{\rm pr} = 0.5 \; (1 + 0.393 \; y^{(0)}) \; N$
($N_{\rm tr} = 0.5 \; (1 - 0.393 \; y^{(0)}) \; N$).   
The overall d$N/$d$y^{(0)}$ distribution was obtained by extrapolating 
the measured transverse momentum spectra, within the backward detector 
acceptance, according to the procedure described in~\cite{hon 98}. 
After deconvolution a shift between the two 
deduced rapidity distributions emerges, demonstrating that a 
memory of the initial target/projectile
translatory motion survives throughout a central collision. 

The separated projectile (or target) rapidity distribution can be readily
parametrized into: (i) a shift  
$\langle y_{\rm {pr}}^{(0)} \rangle$ of the mean value
with respect to mid-rapidity; 
$\langle y_{\rm {pr}}^{(0)} \rangle = 0$ corresponds to
full stopping and full thermo/chemical equilibrium, while
$\langle y_{\rm {pr}}^{(0)} \rangle = 1$ corresponds to 
the initial projectile rapidity without any stopping; positive values 
of $\langle y_{\rm {pr}}^{(0)} \rangle$
are expected for transparency, negative values
for a backward rebound of the projectile nucleons from the target.
(ii) a mixing value ${M}_{\rm pr} = {({N_f - N_b})}/{({N_f + N_b})}$, 
where $N_{f}$ is the number of projectile nucleons emitted forward
and $N_{b}$ \hbox{backwards;} 
${M}_{\rm pr}$ is a measure of non-equilibrium
effects. (iii) a width of the unfolded distribution, ${\sigma}_{\rm pr}$.

\begin{table}[htb]
\caption{
Mean value $\langle y_{\rm {pr}}^{(0)} \rangle$,
width ${\sigma}_{\rm pr}$ of the projectile rapidity
distribution (in $y^{(0)}$ units) and mixing observable
${M}_{\rm pr}$ (see text) obtained for Ru+Ru.
Systematic errors on the values of $\langle y_{\rm {pr}}^{(0)} \rangle$ and
${M}_{\rm pr}$ extracted from the data are about $10 \%$.
(1): IQMD predictions obtained by applying the (${N/Z}$)-tracer method.
(2): same as (1) but using the detector filter.
(3): same as (1) but here IQMD predictions are 
obtained by tagging projectile and target nucleons in the model.
(4): same as 
(3) with $\sigma_{\rm nn}^{\rm med} = 0.5 \; \sigma_{\rm nn}^{\rm free}$.
(5): same as 
(3) with $\sigma_{\rm nn}^{\rm med} = 1.2 \; \sigma_{\rm nn}^{\rm free}$.
(6): same as (3) but without MDI.
}
\begin{center}
\begin{tabular}{cccccccc} 
& & \multicolumn{6}{c}{IQMD} \\ \cline{3-8}
& Data & (1) & (2)  & (3) & (4)  & 
(5) & (6) \\
\hline
$\langle y_{\rm {pr}}^{(0)} \rangle$  & 
0.11 & 0.15 & 0.15 & 0.16  & 0.33  & 0.11  & 0.10  \\  
${\sigma}_{\rm pr}$  & 0.52 & 0.55 & 0.56 & 0.55 & 0.59 & 
 0.54 &  0.52  \\
${M}_{\rm pr}$  & 0.17 & 0.22 & 0.21 & 0.23  &  0.43 & 
 0.16 &  0.15  \\
\end{tabular}
\end{center}
\normalsize
\end{table}

\vspace{-0.3cm}
The defined values are well suited in order to characterize with
few numbers the separated distributions and the strength of the
non-equilibrium effects. 
They are useful in order to verify with the help of theoretical 
models~\cite{har 98,bas 98,hom 99}
the relevance of the (${N/Z}$)-tracer method and effects due to 
the experimental filter. Such studies have been performed here in the 
context of the {\small IQMD}-model~\cite{har 98}. 
Calculations were done using a stiff ($K=380$ MeV) 
{\small EOS} and momentum dependent
interactions ({\small MDI}) for impact parameter selections
similar to those applied to the data.
The results (Table I) show that the extracted values are reliable.
As shown from the {\small IQMD} \hbox{calculations,}
the effect of the detector filter
is negligible. This is due to the fact that the
method relies on relative quantities.
The systematic errors ($\sim 10 \%$) on the experimental
$\langle y_{\rm {pr}}^{(0)} \rangle$ and ${M}_{\rm pr}$ values
take into account the influence of other particle species (t,$\alpha$ and
heavier fragments) not considered in the analysis.     
It is interesting to note also in Table I that the values 
obtained by applying the (${N/Z}$)-tracer method to the 
{\small IQMD} events are
very close to those extracted by tagging the projectile
and target nucleons in the model. This confirms the validity of the
(${N/Z}$)-tracer method proposed in this Letter. 
The dependence of the extracted parameters on the in-medium (n,n)-cross 
section, $\sigma_{\rm nn}^{\rm med}$, is shown in Table I. Both observables
$\langle y_{\rm {pr}}^{(0)} \rangle$ and ${M}_{\rm pr}$ exhibit a quite
strong sensitivity to $\sigma_{\rm nn}^{\rm med}$ 
and depend also on {\small MDI} effects.
On the other hand, the influence of the stiffness of 
the {\small EOS} was found to be very weak: 
calculations using a soft {\small EOS} ($K=200$ MeV) led, within 
statistical uncertainties, to the same results.
A comparison to the experiment \hbox{leads,} within this model, to a value of  
$\sigma_{\rm nn}^{\rm med}$ slightly higher (by about $20 \%$) than 
the free (n,n)-cross section, $\sigma_{\rm nn}^{\rm free}$: a significant 
reduction of $\sigma_{\rm nn}^{\rm med}$ seems to be excluded.

In conclusion, the proposed experiment and method demonstrate
a high sensitivity to non-equilibrium \hbox{effects,} which are found to
persist up to the end of a reaction even in the most central collisions.
This confirms the necessity to use non-equilibrium dynamical
calculations for studies of in-medium effects and of the
{\small EOS} of high density nuclear matter.

This work was supported in part by the French-German agreement between
GSI and IN2P3/CEA, by the PROCOPE-Program of the DAAD and by the
\hbox{Korea} Research Foundation (Contract No. 1997-001-D00117).
The BMBF has funded the collaboration
under the contracts RUM-005-95, POL-119-95 and UNG-021-96.
The DFG  has given support in the framework of the
projects 436 RUS-113/143/2 and 446 KOR-113/76/0.
One of us (Y.L.) would also like to acknowledge support from the
Margarete-von-Wrangell-Program.


\begin{references}
\bibitem{sto 86} H.~St\"{o}cker and W.~Greiner,
         Phys. Rep. {\bf 137} (1986) 277.
\bibitem{ber 84} G.~F.~Bertsch, H.~Kruse and S.~Das~Gupta,
         Phys. Rev.  {\bf C 32} (1984) R673.
\bibitem{aic 91} J.~Aichelin,
         Phys. Rep. {\bf 202} (1991) 233.
\bibitem{fel 90} H.~Feldmeier,
         Nucl. Phys. {\bf A 515} (1990) 147.
\bibitem{hon 98} B.~Hong {\em et al.}, FOPI Collaboration,
         Phys. Rev. {\bf C 57} (1998) 244.
\bibitem{jeo 94} S.~C.~Jeong {\em et al.}, FOPI Collaboration,
         Phys. Rev. Lett. {\bf 72} (1994) 3468.
\bibitem{rei 97} W.~Reisdorf {\em et al.}, FOPI Collaboration,
Nucl. Phys. {\bf A 612} (1997) 493.
\bibitem{yen 94} S.~J.~Yennello {\em et al.},
         Phys. Lett. {\bf B 321} (1994) 15.
\bibitem{joh 96} H.~Johnston {\em et al.}
         Phys. Lett. {\bf B 371} (1996) 186.
\bibitem{joh 97} H.~Johnston {\em et al.}
         Phys. Rev. {\bf C 56} (1997) 1972.
\bibitem{gob 93} A.~Gobbi {\em et al.}, FOPI Collaboration,
         Nucl. Inst. Meth. {\bf A 324} (1993) 156.
\bibitem{rit 95} J.~Ritman  {\em et al.}, FOPI Collaboration,
         Nucl. Phys. (Proc. Suppl.) {\bf B 44} (1995) 708.
\bibitem{lis 95} M.~A.~Lisa, {\small EOS} Collaboration,
             Phys. Rev. Lett. {\bf 75} (1995) 2662.
\bibitem{nag 81} S.~Nagamiya {\em et al.},
         Phys. Rev. {\bf C 24} (1981) 971.
\bibitem{har 98} C.~Hartnack {\em et al.},
         Eur. Phys. J. {\bf A 1}, (1998) 151.
\bibitem{bas 98} S.~A.~Bass {\em et al.},
          Prog. Part. Nucl. Phys. {\bf 41} (1998) 225.
\bibitem{hom 99} A.~Hombach, W.~Cassing and U.~Mosel,
         Eur. Phys. J. {\bf A 5}, (1999) 77.
\end{references}
\end{document}